\documentclass[twocolumn,prx,amsmath,letterpaper,floatfix,showpacs]{revtex4-2}
\usepackage{graphicx}
\usepackage{amsmath}
\usepackage{amssymb}
\usepackage{color}
\usepackage{siunitx}

\newcommand{\otwo}[0]{$\mathrm{O_2}$}
\newcommand{\ntwo}[0]{$\mathrm{N_2}$}

\begin{document}

\title{State- and molecule-selective rotational control in gas mixtures with a shaped optical centrifuge}

\author{P.~Amani, A.~A.~Milner, and V.~Milner}

\affiliation{Department of  Physics \& Astronomy, The University of British Columbia, Vancouver, Canada}

\date{\today}

\begin{abstract}
We demonstrate experimentally a method of all-optical selective rotational control in gas mixtures. Using an optical centrifuge - an intense laser pulse whose linear polarization rotates at an accelerated rate, we simultaneously excite two different molecular species to two different rotational frequencies of choice. The new level of control is achieved by shaping the centrifuge spectrum according to the rotational spectra of the centrifuged molecules. The shaped optical centrifuge releases one molecular species earlier than the other, therefore separating their target rotational frequencies and corresponding rotational states. The technique will expand the utility of rotational control in the studies of the effects of molecular rotation on collisions and chemical reactions.
\end{abstract}

\maketitle

\section{Introduction}
\label{Sec-intorduction}
Intense laser pulses have been long used for controlling molecular rotation (for reviews on this broad topic, see Refs.~\citenum{Stapelfeldt2003,Ohshima2010,Fleischer2012}). The control mechanism is based on the interaction of the electric field of an optical wave with the induced electric dipole of the molecule. The ensuing torque, exerted by the field on the molecule, affects the molecular rotational dynamics \cite{Zon1975, Seideman1995, Friedrich1995, Vrakking1997, Dooley2003, Underwood2005}. The ability to steer these dynamics towards the desired direction and frequency of rotation is key to numerous applications in molecular science \cite{KremsBook}.

When executed with an ultrashort laser pulse (the so-called laser ``kick''), the strength of the instantaneous light-induced torque is a function of the initial orientation of the molecular frame with respect to the polarization of the field. Random orientation angles of molecular axes in a thermal ensemble result in a broad distribution of torque magnitudes and correspondingly broad distribution of final rotational states after a single kick \cite{Averbukh2001, Milner2016a}. Hence, despite the difference in molecular properties, a single non-resonant pulse typically generates overlapping rotational distributions in different molecular species with limited selectivity between them.

To achieve molecular selectivity, sequences of two or three properly timed laser pulses have been used \cite{Renard2004, Lee2006, Fleischer2006, Fleischer2007}. The desired selectivity stems from the quantum interference between the rotational wave packets created by multiple kicks, which allows one to increase the amplitude of one quantum state, while decreasing the other. Though successful in selective rotational excitation of molecular isotopes \cite{Fleischer2006} or spin isomers \cite{Fleischer2007} in a mixture, the method does not offer a means of controlling the target rotational states in the excited molecules. Increasing the number of pulses promised to add the state selectivity \cite{Zhdanovich2012}, but has not proved practical due to the technical challenges of creating long pulse trains on a picosecond time scale \cite{Bitter2016b}.
\begin{figure*}[t]
    \includegraphics[width=2\columnwidth]{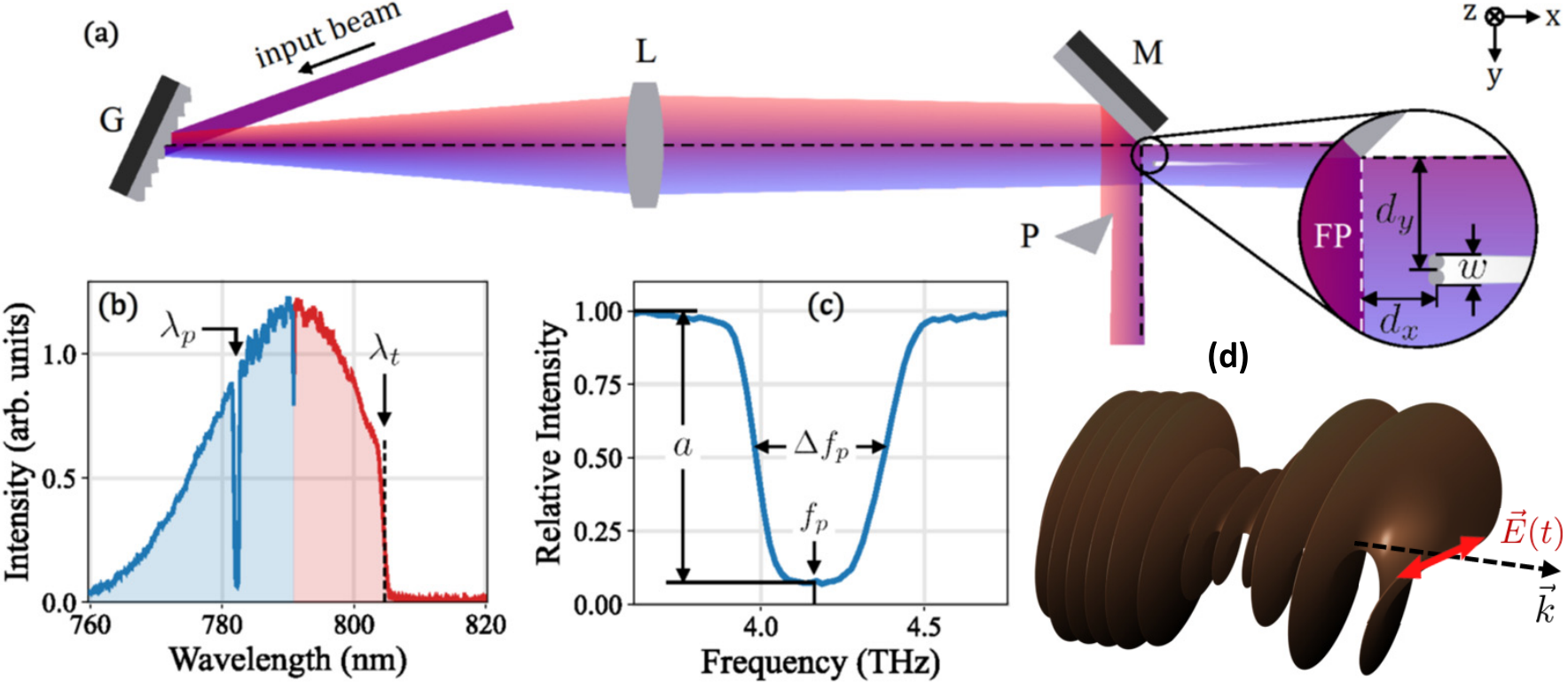}
    \caption{Illustration of the main principle behind shaping the field of an optical centrifuge with the ``truncation'' and ``piercing'' techniques.
    (\textbf{a}) Optical setup of the input segment of the centrifuge shaper. G: diffraction grating, L: lens, M: mirror, P: truncating prism, FP: Fourier plane. The inset shows the use of optical fibers (two grey circles) for blocking a narrow segment in the centrifuge spectrum. The location ($d_x, d_y, d_z$ - see the coordinate axes at the upper right corner) and width ($w$) of the fibers determine the properties of the created frequency notch.
    (\textbf{b}) \textit{Optical} spectrum of the shaped centrifuge, showing both the truncation of the red arm (red shaded area) at $\lambda_t=805$~nm, and the piercing  of the blue arm (blue shaded area) at $\lambda_p=783$~nm.
    (\textbf{c}) Frequency notch in the \textit{rotational} spectrum of the pierced centrifuge (see text for the definition of the term). $f_p$ and $\Delta f_p$ describe the rotational frequency window, within which the centrifuge intensity is reduced by a factor $(1-a)$, where $a$ is the piercing depth.
    (\textbf{d}) Calculated polarization profile of the pierced centrifuged. The vector of linear polarization $\vec{E}(t)$ undergoes accelerated rotation around the wave vector $\vec{k}$. The lower-amplitude ``neck'' in the middle of the pulse is a result of the spectral piercing. }
    \label{Fig-shaper}
\end{figure*}

A powerful method of controlling the target rotational state of a molecule has emerged in 1999, when an ``optical centrifuge'' (OC) has been theoretically proposed \cite{Karczmarek1999} and later experimentally demonstrated \cite{Villeneuve2000}. An optical centrifuge is a laser pulse, whose polarization vector undergoes an accelerated rotation around the propagation direction. An interaction of the induced dipole moment with such a field forces molecules to follow the polarization vector, much like an object placed inside a mechanical centrifuge follows its rotary motion. Quantum mechanically, the process can be described as a series of adiabatic transitions, through which a molecule climbs the ladder of rotational energy levels \cite{Spanner2001, Spanner2001a, Armon2017}. In contrast to the non-adiabatic kicks, OC is capable of creating a narrow rotational wave packet centered at a well-defined target rotational state \cite{Korobenko2014a}.

To date, optical centrifuges entered numerous studies (for a recent review, see Ref.~\citenum{MacPhail2020}). Yet despite the analogy implied by its name, the optical centrifuge has never been used to select molecules according to their physical properties and control their rotation separately (similarly to the selectivity of its mechanical counterpart, based on the material density). Rather, the centrifuge has always defined  a single rotational frequency in any given experiment. Investigations of the effects of rotational excitation on the collisional properties \cite{Yuan2011, Toro2013, Korobenko2014a, Milner2014a, Murray2018}, gyroscopic dynamics \cite{Milner2015c, Murray2017} or chemical reactions \cite{Ogden2019} have all been limited to having either both collision partners rotating with the same frequency, or one of them being a randomly rotating molecule at equilibrium with the thermal ensemble. This rather limiting constraint stems from the binary ``all or nothing'' selective power of the centrifuge: it either overcomes the thermal motion of the molecules and spins them up, or leaves them behind if the interaction potential is too weak or the angular acceleration is too fast.

In this work, we demonstrate a method of adding molecular selectivity to the rotational control with a \textit{shaped optical centrifuge}. By modifying the spectrum of the centrifuge, we make it spin different molecular species in a gas mixture to different angular frequencies. Both the separation and the controllability of target frequencies is accomplished by removing certain segments from the OC spectrum (hereafter referred to as spectral ``piercing'') according to the rotational spectra of the molecules. By blocking a resonant Raman line in the rotational ladder of one molecule, we force it to stop climbing the ladder midway. At the same time, another molecule keeps climbing to higher frequencies, as long as the missing spectral component does not belong to the rotational ladder of that molecule.
\begin{figure*}[t]
    \includegraphics[width=1.95\columnwidth]{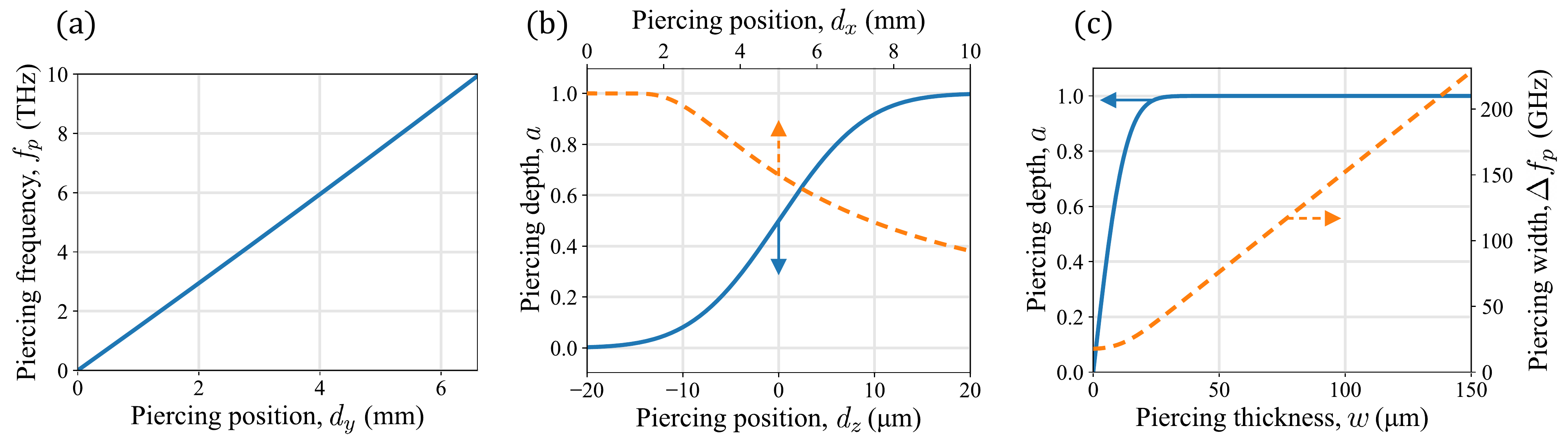}
    \caption{Numerically calculated characteristics of the frequency notch, applied to the rotational spectrum of the centrifuge by means of the spectral piercing method and depicted in Fig.~\ref{Fig-shaper}(\textbf{c}) . (\textbf{a}) Frequency of the OC rotation, blocked by piercing and defined by the position of the piercing element in the Fourier plane of the shaper, $d_y$. (\textbf{b}) Shifting the piercing element along the two other directions, $d_z$ (at $d_x=0$, solid line) or $d_x$ (at $d_z>\SI{20}{\micro\meter}$, dashed line), controls the depth of the frequency notch $a$. (\textbf{c}) The thickness of the piercing element $w$ defines the width of the frequency notch, $\Delta f_p$ (dashed line), but also affects its depth $a$ (solid line).}
    \label{Fig-piercing}
\end{figure*}

\section{Spectral shaping of an optical centrifuge}
\label{Sec-shaper}
To create the field of an optical centrifuge, the spectrum of a broadband laser pulse is split in two equal parts (hereafter referred to as ``centrifuge arms'') using a Fourier pulse shaper \cite{Villeneuve2000}. The two arms are frequency chirped with opposite chirps, circularly polarized with opposite handedness, and overlapped in space and time. Interference of the two circularly polarized laser fields results in a rotating linear polarization, whereas the increasing frequency difference between the two centrifuge arms due to the opposite frequency chirps makes the latter rotation accelerate with time (for specific sets of parameters, see Ref.~\citenum{MacPhail2020}).

We use a regenerative Ti:Sapphire amplifier, which produces laser pulses with 10~mJ energy per pulse and 35~fs pulse length at a central wavelength of 790~nm and a repetition rate of 1~KHz. Fig.~\ref{Fig-shaper}(\textbf{a}) schematically depicts the input segment of our centrifuge shaper (for the full optical layout, which replicates the original design of Villeneuve \textit{et al.} \cite{Villeneuve2000}, see for example Fig.~3 in \cite{MacPhail2020}). Mirror M splits the spectrum of the centrifuge in the Fourier plane (FP) of Lens L. Using the standard combination of lenses and diffraction gratings (not shown), a negative frequency chirp is applied to the reflected red arm, whereas the transmitted blue arm is chirped positively.

In addition to the frequency chirping, introduced by the conventional OC shaper, we modify the centrifuge spectrum as follows. An array of glass fibers is introduced in the Fourier plane in order to block (by means of light scattering) a small part of the blue arm's spectrum, as shown in the inset to Fig.~\ref{Fig-shaper}(\textbf{a}). Glass fibers of optical quality are chosen for the purpose of withstanding high light intensities, with up to 10 fibers (125~$\mu$m diameter each) grouped together in a tight side-by-side geometry. The spectrum of the red arm is truncated by a glass prism, positioned close to the Fourier plane and blocking the red spectral tail of the centrifuge (again, by scattering that light out of the shaper).

The effects of both the piercing of the blue arm and the truncation of the red arm are demonstrated in the OC spectrum plotted in Fig.~\ref{Fig-shaper}(\textbf{b}). To understand the effect of these modifications in the \textit{optical} spectrum of the centrifuge field, we recall the expression for the two centrifuge arms in the time domain \cite{MacPhail2020}:
\begin{equation}\label{Eq-onePhoton}
\vec{E}_\pm(t)=\frac{E_0(t)}{2} \hat{\epsilon}_\pm e^{-i(\omega_0 \pm \beta t)t,}
\end{equation}
\noindent where $E_0(t)$ and $\omega _0$ are the field envelope and the central frequency of the OC pulse, $\hat{\epsilon}_\pm$ are the unit vectors of two circular polarizations [with `+' (`-') being the polarization of the blue (red) arm], and $\beta $ is the angular acceleration of the centrifuge. The instantaneous optical frequencies of these fields are $\omega (t) = \omega _0\pm 2\beta  t$. Hence, piercing the spectrum of the blue arm $\vec{E}_+ (t)$ at frequency $\omega _p$ corresponds to attenuating the field at time $t_{p} = (\omega _{p}-\omega _0)/2\beta $. Similarly, truncating the red arm at $\omega _t$ results in its termination at time $t_{t} = (\omega _0 -\omega _t)/2\beta $. This is illustrated in Fig.~\ref{Fig-shaper}(\textbf{d}), which shows the numerically calculated centrifuge pulse, propagating in the direction $\vec{k}$, with its vector of linear polarization $\vec{E}(t)$ rotating at an accelerated rate around the propagation axis. Piercing of the blue arm results in a temporary drop of the field amplitude in the middle of the OC pulse, whereas the truncation of the red arm stops the accelerated rotation of the polarization vector. Note that the circularly polarized fields of both the non-pierced red arm in the middle section, and the non-truncated blue arm at the end of the pulse, are not shown in Fig.\ref{Fig-shaper}~(\textbf{d}) for clarity. Being far off any electronic, vibrational and rotational resonances, this light has no significant effect on the molecular dynamics.

The rotational ladder climbing is a sequence of two-photon stimulated Raman transitions, during which a molecule absorbs one photon from the blue arm and emits one photon into the red arm. The process is therefore governed by the two-photon difference-frequency-generation (DFG) spectrum of the OC (hereafter referred to as the ``rotational'' spectrum of the centrifuge). The two-photon field in the time domain is a product of the two centrifuge arms:
\begin{equation}\label{Eq-twoPhoton}
E^{(2)}_\text{DFG}(t):= \vec{E}_+(t) \cdot \vec{E}^*_-(t) = \frac{E_0^2(t)}{2} e^{-i 2\beta t^2}.
\end{equation}
\noindent The instantaneous frequency of this field is $\Omega (t) = 4\beta t$. By substituting $t_{p,t}$ from the previous paragraph, one finds that piercing (truncating) the centrifuge spectrum at $\omega _p$ ($\omega _t$) results in the attenuation (termination) of the rotational ladder climbing at $\Omega _{p,t}=\left| 2(\omega _{p,t} - \omega _0) \right|$.

An example of the spectral notch in the two-photon spectrum of the pierced centrifuge is shown in panel (\textbf{c}) of Fig.~\ref{Fig-shaper}. The notch is characterized by the central frequency $f_p$, width $\Delta f_p$ (full width at half maximum), and depth $a$. The relationship between these quantities and the position/dimensions of the piercing element $d_x, d_y, d_z$ and $w$, indicated in the inset of Fig.~\ref{Fig-shaper}(\textbf{a}), can be found by using the known rules of Fourier optics \cite{GoodmanBook}. Numerically calculated $f_p, \Delta f_p$ and $a$ for the parameters of our centrifuge shaper (groove density of 1500 mm$^{-1}$ and focal length of 20~cm for the diffraction grating G and the focusing lens L, respectively) are shown in Fig.~\ref{Fig-piercing}.

Panel (\textbf{a}) of Fig.~\ref{Fig-piercing} shows the linear transformation of the rotational centrifuge spectrum to the spatial distribution in the Fourier plane, similar to any standard `$4f$' pulse shaper \cite{Weiner2000}. By moving the piercing element on the scale of a few millimeters, the notch can be introduced at any Raman frequency between 0 and 10~THz. Note that the same frequency-to-space conversion of $\approx 1.5$~THz/mm applies to the truncation of the OC spectrum with prism P, as discussed above.

Fig.~\ref{Fig-piercing}(\textbf{b}) illustrates the two possible mechanisms for controlling the piercing depth $a$. It can be executed by either moving the piercing element (here, a glass fiber of $\SI{125}{\micro\meter}$ diameter) in a perpendicular direction with respect to the dispersion plane of the centrifuge shaper ($d_z$), or by moving it away from the Fourier plane ($d_x$). In the former case, the length scale is defined by the diffraction-limited beam radius of a monochromatic beam, $w_0=\SI{15}{\micro\meter}$ at $d_x=0$. Increasing $d_x$ on the scale of the Rayleigh length, $x_R=\SI{0.4}{\milli\meter}$, with the fiber fully inserted along the z-axis ($d_z > w_0$), governs the second mechanism of attenuating a particular range of Raman frequencies. The latter approach may be useful for controlling the steepness of the frequency notch, e.g. for increasing the adiabaticity of the molecule-centrifuge interaction at the piercing frequency.

The ability to control the width of the frequency notch in the rotational spectrum of the OC by varying the thickness of the piercing element ($w$) is demonstrated in panel (\textbf{c}) of Fig.~\ref{Fig-piercing}. As expected, the notch width grows linearly with $w$, as long as the latter is bigger than $\SI{20}{\micro\meter}$, which is defined by the diffraction limit along the $\hat{y}$-axis. The diffraction limit also determines the narrowest piercing, which in the case of our shaper amounts to $\Delta f_p= \SI{18}{\giga\hertz}$. Note, however, that approaching that minimum comes at the cost of incomplete attenuation, $a<1$. An incomplete attenuation may also stem from a small leakage of light through the piercing element. Glass fibers, for example, withstand high intensities in the Fourier plane, but limit the piercing depth to $a\lesssim 90\%$. This constraint is not included in Fig.~\ref{Fig-piercing}, but was encountered in our experiments discussed below.

Our experimental setup for the state-selective excitation of molecules to high rotational states (also known as ``super-rotors'' states) has been described in previous publications and summarized in Ref.~\citenum{MacPhail2020}. Briefly, the centrifuge pulses are focused in the cell filled with a gas of interest at room temperature and variable pressure. A positive lens with the focal length of 10~cm provides the length of the centrifuged region of about 1~mm and peak intensities of up to $5\times 10^{12}$~W/cm$^{2}$. Higher intensities are avoided due to the detrimental strong-field effects, such as multi-photon ionization and filamentation.

For the state-resolved detection of optically centrifuged molecules, we use polarization-sensitive rotational Raman spectroscopy. Each centrifuge pulse is followed by a weak circularly polarized probe pulse, derived from the same laser system and spectrally narrowed down to the bandwidth of $0.1$~nm (pulse length of $\sim3$~ps). Frequency doubling of probe pulses shifts their central wavelength to 395~nm, which allows an easy separation from the centrifuge beam. Coherent forward scattering of the probe light by an ensemble of centrifuged molecules results in a rotational Raman shift, whose magnitude is equal to twice the rotational frequency, whereas the sign indicates the direction of molecular rotation with respect to the circular probe polarization \cite{Korech2013, Korobenko2014a}.

\begin{figure}[t]
    \includegraphics[width=0.83\columnwidth]{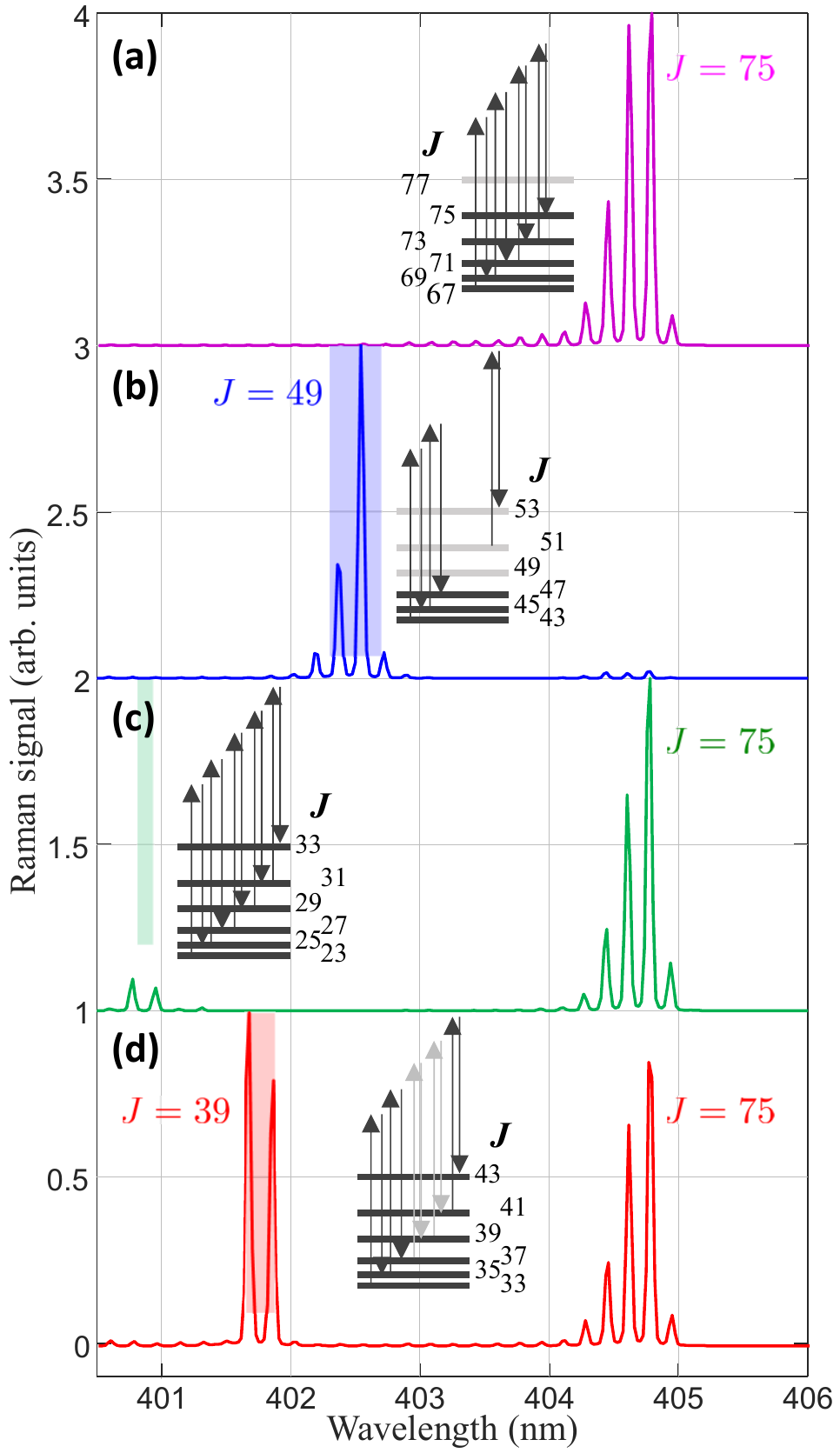}
    \caption{Illustration of the main concept of spectral ``piercing'' of an optical centrifuge. Raman signals in a gas of \otwo{} molecules at room temperature and pressure of 36~kPa. From top to bottom, the spectra were obtained with (\textbf{a}) a centrifuge truncated at 6~THz;  (\textbf{b}) a hard pierced centrifuge for stopping the rotational acceleration around 4~THz; (\textbf{c}) a soft pierced truncated centrifuge for letting the molecules accelerate to 6~THz; (\textbf{d}) a medium pierced truncated centrifuge for splitting the ensemble between the two rotational frequencies of 3.5 and 6~THz. The strongest line in each rotational wave packet is labeled with the corresponding value of the rotational quantum number $J$. Colored rectangles indicate the position, width and depth of the spectral notch in the pierced centrifuge spectrum. The insets illustrate a few consecutive Raman transitions between the rotational energy levels (horizontal lines) near the truncation frequency (\textbf{a}) and the frequency of the notch (\textbf{b-d}). Grey horizontal lines represent (almost) empty levels, while grey vertical arrows denote weaker transitions due to the lower field amplitudes.}
    \label{Fig-o2}
\end{figure}
\section{Control of molecular rotation with a pierced centrifuge}
\label{Sec-o2}
Figure~\ref{Fig-o2}(\textbf{a}) shows an example of the Raman spectrum of oxygen gas, obtained with an unshaped optical centrifuge. The set of discrete Raman lines around 404.5~nm indicates a coherent rotational wave packet, which corresponds to the oxygen super-rotors spinning at a frequency of $\approx 6$~THz in the same direction as the circular polarization of the probe (hence, the frequency down-shifted Stokes lines). The wave packet consists mostly of three eigenstates with the rotational quantum numbers $J=71, 73$ and 75 \footnote{Note that in the case of \otwo{} molecules, $J$ should be understood as an average value between the three unresolved spin-rotational components $J=\{N,N\pm 1\}$, where $N$ is the nuclear rotation quantum number.}.

As we have demonstrated in a number of our previous works (e.g., see Ref.~\citenum{Milner2016a}), truncating the centrifuge spectrum, as illustrated in Fig.~\ref{Fig-shaper}, enables us to control the target rotational state of the super-rotors. For a given angular acceleration of the OC (here, 0.3~rad/ps$^{2}$), the rotational state of the centrifuged molecules is determined by the time of their interaction with the centrifuge field. The implemented spectral truncation shortens the duration of the pulse and, therefore, the corresponding interaction time, thus lowering the molecular final rotational frequency. In the language of consecutive Raman transitions, the rotational ladder climbing starts from the initial state and continues uninterrupted until the highest possible step, dictated by the truncated edge of the OC spectrum, as depicted in the inset to panel (\textbf{a}) in Fig.~\ref{Fig-o2}.

Fig.~\ref{Fig-o2}(\textbf{b}) illustrates an alternative way of rotational control, based on stopping the rotational acceleration of molecules prematurely, i.e. before the end of the centrifuge pulse. Piercing the centrifuge spectrum with a relatively deep notch filter, which blocks two consecutive Raman transitions, interrupts the rotational excitation for just enough time to make it stall. In the example of Fig.~\ref{Fig-o2}(\textbf{b}), the notch (schematically shown with a blue rectangle) takes out two steps from the rotational ladder, $J=47\rightarrow J=49$ and $J=49\rightarrow J=51$ (note missing Raman transitions in the inset). With those two steps absent, the molecules pile up at the rotational states with $J=47, 49$. This approach may be advantageous in those cases when it is important to maintain constant pulse energy while controlling the target rotational frequency. In comparison to the spectral truncation described above and typically leading to large energy changes (often exceeding 50\%), spectral piercing reduces the pulse energy by as little as 2\%. An example of utilizing this property of the pierced centrifuge can be found in our recent work on the detection of the mechanical Faraday effect in gaseous media \cite{Milner2021a}.

In contrast to the blocking action of the spectral notch covering two Raman transitions, a narrower hole in the spectrum may have little effect on the centrifuge excitation if it is placed between two consecutive Raman lines. An example is shown in Fig.~\ref{Fig-o2}(\textbf{c}), where the spectrum is pierced between the transitions $J=29\rightarrow J=31$ and $J=31\rightarrow J=33$. Owing to the limited resolution of our pulse shaper, the notch is not sufficiently narrow to let all oxygen molecules pass through without causing some of them to fall out of the centrifuge. This is reflected by a small Raman signal originating from the two states with $J=29, 31$. Yet the majority of the molecules, caught by the centrifuge, continue their rotational acceleration and reach the same target states around $J=75$ as in the case of a non-pierced centrifuge. Both the amplitude and the shape of the final rotational wave packet are hardly changed by the piercing procedure.

By partially overlapping the spectral notch in the centrifuge spectrum with one or two Raman transitions, one can split the rotational wave packet between two central frequencies in a controlled way. This is illustrated in Fig.~\ref{Fig-o2}(\textbf{d}), where piercing the spectrum so as to partially cover the $J=39\rightarrow J=41$ and $J=41\rightarrow J=43$ Raman lines, results in two equally-populated wave packets. We note that the relative amplitudes of the two wave packets are fully controllable by varying the depth of the spectral notch (here, adjusted for a $\approx$~50/50 split). Together with the freedom in choosing both the piercing and the truncation wavelengths, this gives us complete rotational control over the two groups of molecules.

\section{Selective spinning of molecules in mixtures}
\label{Sec-2molecules}
As one can see from the described examples in Fig.~\ref{Fig-o2}, the effect of the centrifuge piercing on a particular molecule depends on the relative position of the hole in the OC spectrum with respect to the Raman resonances of that molecule. This suggests that a centrifuge may be pierced in such a way as to stop the acceleration of one molecular species at a lower frequency, defined by the location of the spectral notch, while spinning the other one higher up.

\begin{figure}[t]
    \includegraphics[width=0.99\columnwidth]{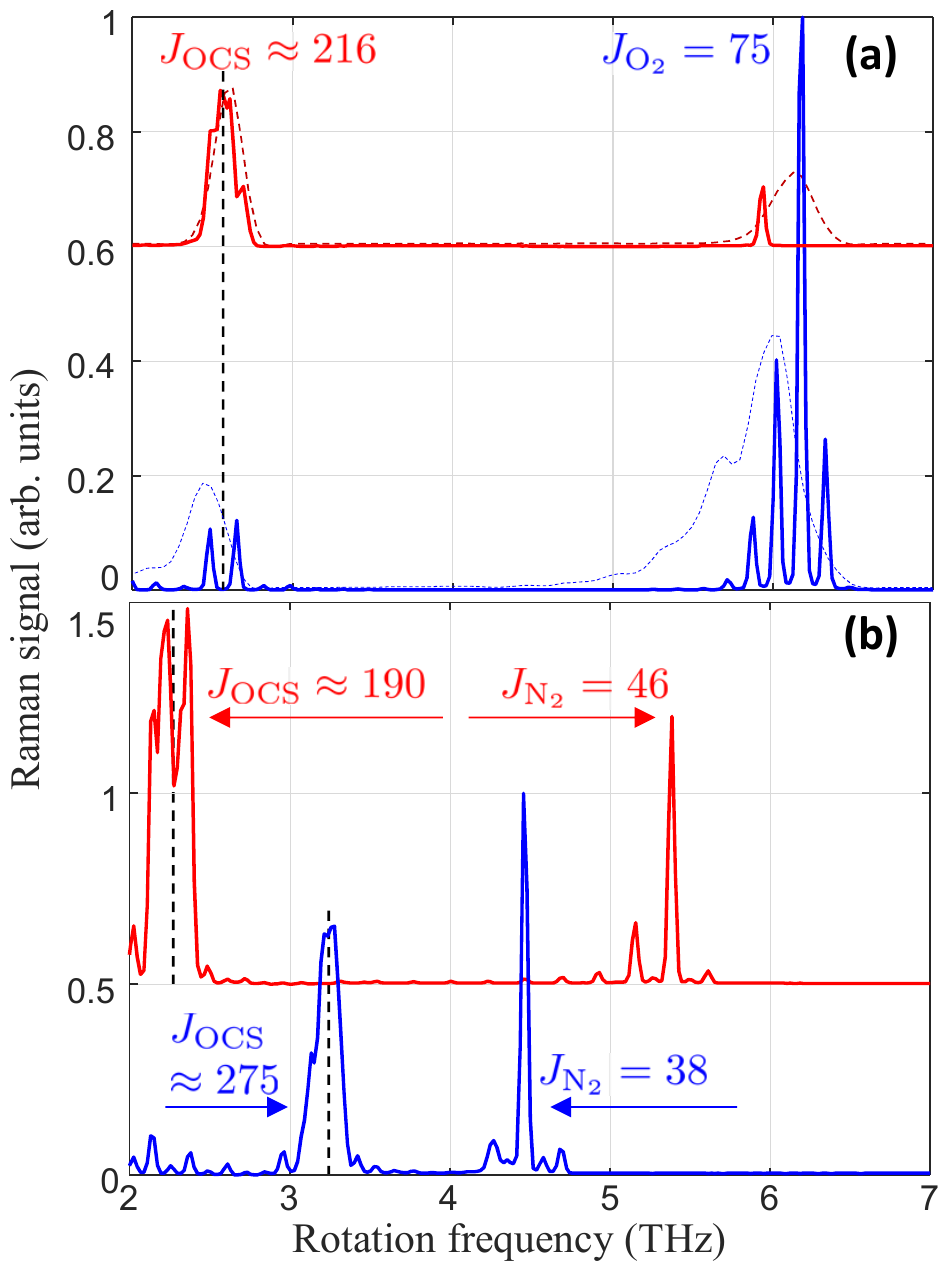}
    \caption{Selective rotational excitation by a pierced optical centrifuge in (\textbf{a}) a gas sample of either OCS or \otwo{} molecules at room temperature and pressure of 10~kPa and 36~kPa, respectively, and (\textbf{b}) a mixture of OCS with \ntwo{} gases at room temperature and partial pressure of 7~kPa and 30~kPa, respectively. All lines represent rotational Raman spectra, collected with the probe pulses arriving between 40~ps and 50~ps after the end of the centrifuge pulse, and integrated over that time window. The labels indicate the rotational quantum number of the most populated rotational state in the corresponding wave packet. Vertical dashed lines depict the central frequency of the spectral notch, introduced by the piercing procedure. The dashed lines in panel (\textbf{a}) show the results of numerical simulations (see text for details).}
    \label{Fig-2molecules}
\end{figure}

Fig.~\ref{Fig-2molecules}(\textbf{a}) shows an example of applying the pierced centrifuge separately to the gas of OCS and \otwo{} molecules. Due to the much higher moment of inertia, the rotational spectrum of carbonyl sulfide is seven times denser than oxygen's spectrum. This means that even the narrowest piercing available with our pulse shaper will cover quite a few Raman transitions in the OCS spectrum, making the notch largely ``impassable''. The top red curve in Fig.~\ref{Fig-2molecules}(\textbf{a}) confirms that the majority of carbonyl sulfide molecules ended their angular acceleration at the location of the spectral notch around 2.5~THz, which corresponds to the most populated rotational state with $J_\text{OCS}\approx216$.

The bottom blue curve in Fig.~\ref{Fig-2molecules}(\textbf{a}) shows that the very same centrifuge, which ``dropped'' OCS around 2.5~THz, spins the majority of oxygen molecules to the high super-rotor states centered at 6.2~THz (or the rotational quantum number $J_{\text{O}_2}=75$). As illustrated in Fig.~\ref{Fig-o2}(\textbf{c}), the high rotational excitation of \otwo{} is accomplished by piercing the centrifuge not too deep and between two Raman transitions, thus making it less disruptive for the rotational ladder climbing. A small leakage of OCS towards higher rotational frequencies, and a small amount of \otwo{} lost at lower frequencies, indicate the incomplete rotational selectivity afforded by the presented piercing technique. The limitation stems from the simplicity of the employed spectral shape, and will be discussed in Sec.~\ref{Sec-summary}.

Our numerical simulations of the molecular spinning in the pierced OC qualitatively reproduce the experimental observations. The simulations are based on solving the system of classical coupled Euler equations and quaternion equations of motion \cite{Tutunnikov2018}. The results are plotted with dashed lines in Fig.~\ref{Fig-2molecules}(\textbf{a}). Interestingly, the numerically obtained contrast in the rotational selectivity between \otwo{} and OCS is considerably smaller than what we achieve in the experiment. The reason for the disagreement is the inability of the classical model to account for the discreteness of the rotational spectrum of a quantum rotor, which lies at the heart of our method.

Selective rotational excitation of two molecular species, mixed together in the same gas cell, is demonstrated in Fig.~\ref{Fig-2molecules}(\textbf{b}). Here, OCS and \ntwo{} were mixed at partial pressures of 7~kPa and 30~kPa, respectively. Similar to the previous example, big difference in the density of rotational states between the two molecules enables one to stop the angular acceleration of OCS earlier than the acceleration of \ntwo{}, which proceeds all the way to the end of the centrifuge pulse. This can be recognized by the presence of two peaks in the Raman spectrum of the mixture - the broad one at a lower frequency and the narrow one at a higher frequency. The peaks can be respectively assigned to carbonyl sulfide, whose lower rotational constant does not allow us to resolve individual lines (hence, larger peak width), and nitrogen, whose Raman resonances are well resolved with our spectrometer. Sharp lines on top of the broad peak at 2.3~THz are due to the small amount of \ntwo{} molecules, which fell out of the pierced centrifuge together with the majority of OCS.

The two Raman spectra in Fig.~\ref{Fig-2molecules}(\textbf{b}) illustrate our ability to control both the slow carbonyl sulfide and the fast nitrogen rotors, independently from one another. By moving the position of the hole in the spectrum of the centrifuge from 787.2~nm to 785.1~nm [$\lambda_p$ in Fig.~\ref{Fig-shaper}(\textbf{b})], we prolonged the life time of OCS in the centrifuge from 24~ps to 35~ps, thus increasing its final rotational frequency from 2.3~THz to 3.3~THz. Similarly, shortening the centrifuge pulse from 57~ps to 47~ps by means of varying the truncation wavelength [$\lambda_t$ in Fig.~\ref{Fig-shaper}(\textbf{b})] slows down the rotation of \ntwo{} from 5.4~THz to 4.5~THz.

\begin{figure}[t]
    \includegraphics[width=0.99\columnwidth]{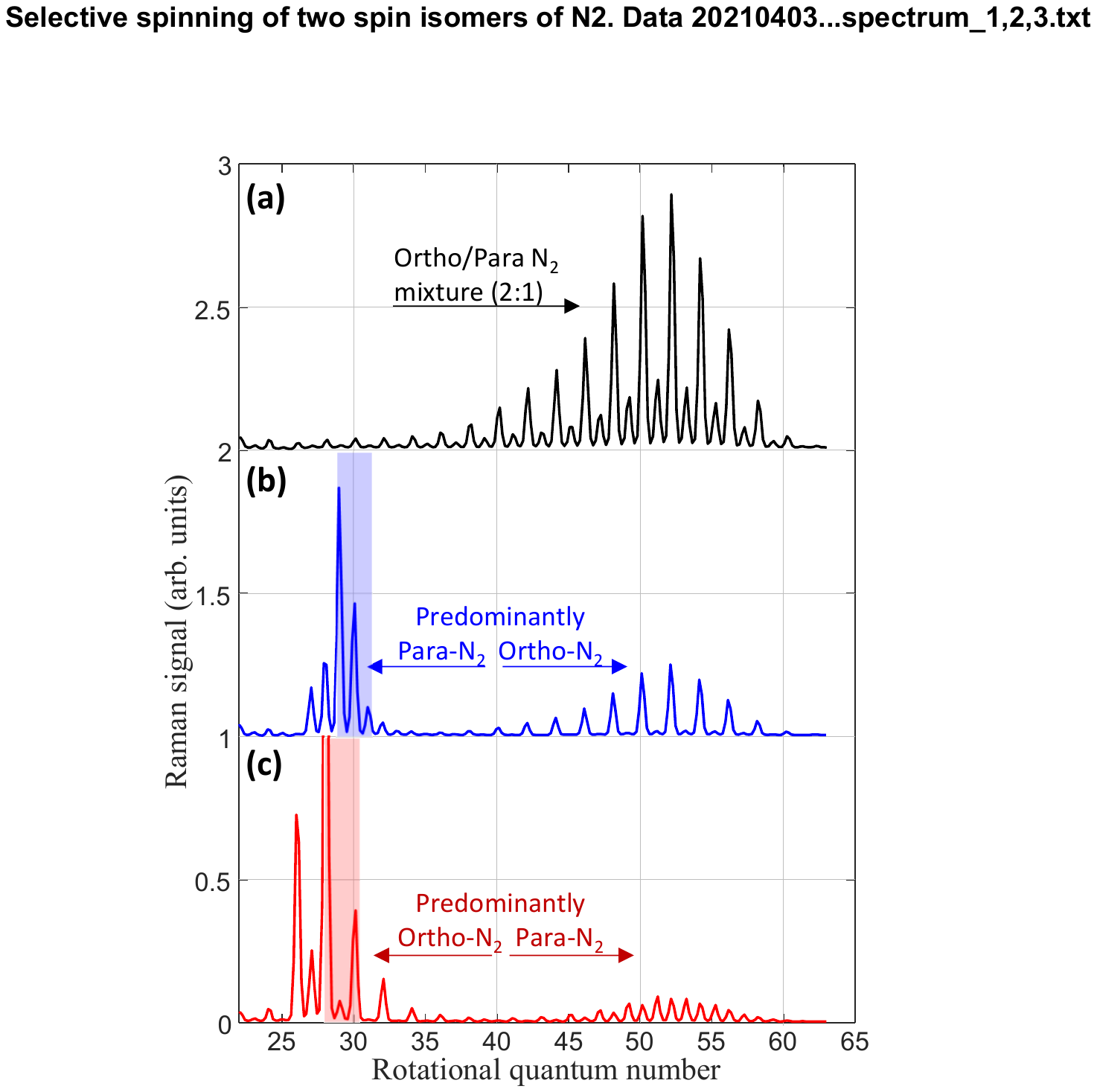}
    \caption{Selective rotational excitation in the mixture of two nuclear-spin isomers of $^{14}\text{N}_2$. Raman spectra are plotted along the rotational quantum number of nitrogen, thus showing the constituent quantum states of each rotational wave packet. The three spectra correspond to (\textbf{a}) a non-pierced optical centrifuge, (\textbf{b}) a centrifuge pierced for decreasing (increasing) the amount of para-nitrogen in the fast (slow) wave packet, and (\textbf{c}) a centrifuge pierced for decreasing (increasing) the amount of ortho-nitrogen in the fast (slow) wave packet. Colored rectangles indicate the position and width of the spectral notch in the pierced centrifuge spectrum.}
    \label{Fig-2isomers}
\end{figure}
\section{Selective rotational excitation of two spin isomers}
\label{Sec-2isomers}
In addition to the rotational selectivity between different molecular species, discussed above, centrifuge piercing can also be utilized in the selective spinning of molecular isotopes and spin isomers. The former would rely on the difference between the rotational constants of the two isotopes \cite{Fleischer2006}, whereas the latter could be based on the different parity constraints for the rotational wave functions of the two spin isomers \cite{Fleischer2007}. Incidentally, for a diatomic homonuclear molecule whose nuclei are bosons, such as $^{14}\text{N}_2$, the symmetric nuclear wave function (known as ortho-nitrogen) must be combined with the symmetric rotational wave function. Hence, ortho-nitrogen has only even $J$ numbers in its rotational spectrum. Similarly, the antisymmetric nuclear wave function of para-nitrogen limits its rotational quantum numbers to odd values only.

Because of the $\Delta J=\pm2$ selection rule, rotational Raman transitions do not couple quantum states of different parity. Hence, by piercing an optical centrifuge in such a way as to suppress the rotational ladder climbing via either even or odd $J$-states, the acceleration of one spin isomer can be terminated early, while the other isomer is accelerated to higher frequency by the end of the OC pulse.

Both spin isomers coexist in an ambient gas of $^{14}\text{N}_2$ with the ortho:para ratio of 2:1. This is reflected by the coherent Raman spectrum of \ntwo{} super-rotors in Fig.~\ref{Fig-2isomers}(\textbf{a}), where an approximately quadratic dependence of the Raman signal on the molecular population \cite{Bitter2016c} results in a $\sim$~4:1 ratio of the peak intensities originating from the two spin isomers. Ortho-nitrogen is excited to a broad wave packet centered at $\bar{J}_\text{ortho}=52$ (the comb of tall lines), whereas the most occupied state of para-nitrogen is $\bar{J}_\text{para}=51$ (the comb of short lines).

Figs.~\ref{Fig-2isomers}(\textbf{b,c}) demonstrate the ability of the pierced OC to differentiate between ortho- and para-nitrogen in terms of their respective rotational excitations. In panel~(\textbf{b}), the location of the notch in the centrifuge spectrum results in blocking two steps in the rotational ladder of para-nitrogen, $J=29\rightarrow J=31$ and $J=31\rightarrow J=33$, and only one step in the climbing path of the ortho-isomer, $J=30\rightarrow J=32$ (depicted by the blue rectangle). As can be seen from the relative line strengths, this piercing largely eliminates para-nitrogen from the fast wave packet (which remains centered at $J=52$), while making it the dominant component of the slow one (centered at $J=29$).

Reversal of the slow--fast separation between the isomers is not as visible in Fig.~\ref{Fig-2isomers}(\textbf{c}), due to the initial prevalence of ortho-nitrogen in the sample. Here, para-nitrogen is slightly dominant at higher $J$'s (wave packet centered at $J=51$), whereas most of ortho-nitrogen has been ``dropped'' by the centrifuge around $J=28$. The separation has again been accomplished by blocking two steps in the ladder of even rotational states, while taking out only a single step from the ladder of odd $J$'s. Note that similarly to the earlier example with the OCS--\ntwo{} mixture, controlling the center frequency of both the slow and fast wave packets can be easily achieved by moving the spectral notch and the outer truncation edge in the centrifuge spectrum, respectively.

\section{Summary}
\label{Sec-summary}
In summary, we developed an experimental method of selective rotational control in gas mixtures. Using a shaped optical centrifuge, we simultaneously excite two different molecular species to two different rotational frequencies of choice. The control is accomplished by piercing the centrifuge spectrum with a notch filter, and aligning its spectral position with respect to the Raman resonances of both molecules. The position of the frequency notch determines whether the molecule falls out of the centrifuge, and if so - with what rotational frequency, or continues climbing the rotation ladder to a higher level of rotational excitation.

As a proof of principle, we demonstrated the selectivity of the pierced centrifuge in a mixture of two different gases, and a mixture of two spin isomers of the same molecule. In the first case, the difference in the density of rotational states allows one to terminate the rotational acceleration of a molecule with the higher moment of inertia earlier than a molecule, whose moment of inertia is lower. As a result, the former is excited to the lower frequency than the latter, with both frequencies being controlled by the shape of the centrifuge spectrum. In the case of spin isomers, the control is accomplished by aligning the spectral notch so as to eliminate more Raman resonances of either ortho- or para-isomer, thus interrupting the acceleration of that isomer earlier than its nuclear-spin counterpart. The second example showcases our ability to differentiate between two molecules with similar (or even equal) moments of inertia, as long as their Raman spectra are sufficiently distinct.

The reported method of rotational control exhibits limited degree of selectivity, in that the rotational wave packets created by the pierced OC are slightly cross-``contaminated'' by the molecules from the other group, as can be seen in Figs.~\ref{Fig-2molecules} and \ref{Fig-2isomers}. We attribute this shortcoming to the simplicity of the applied spectral shaping, which in its current implementation consists of a single frequency notch. Our experiments with OCS and \otwo{} show that making the notch narrower increases the amount of (undesired) fast rotating carbonyl sulfide, whereas widening it causes bigger accumulation of slower oxygen rotors. Neither change in the parameters of the single-notch piercing (including its depth and steepness) results in a better simultaneous suppression of those two groups of molecules, and hence in the improved rotational selectivity. The results of our numerical simulations reflect the same limitation. We therefore conclude, that piercing the centrifuge at a single frequency may not provide complete selectivity, although the latter improves as the difference between the rotational properties of the two molecules increases. Studies are underway to eliminate this constraint by applying more elaborate pulse shapes, e.g., by means of increasing the number of holes in the OC spectrum.

The demonstrated effect adds yet another ``control knob'' to the existing toolbox for harnessing molecular dynamics with laser fields. It can be instrumental in any studies involving molecular collisions or controlled chemical reactions, because of the added ability to control both the absolute and the relative rotational frequencies of the collision/reaction partners.

\section*{ACKNOWLEDGMENTS}
This work was carried out under the auspices of the Canadian Center for Chirality Research on Origins and Separation (CHIROS).


%

\end{document}